\documentclass[twocolumn,showpacs,preprintnumbers,amsmath,amssymb,floatfix]{revtex4}
\usepackage{graphicx}% Include figure files
\usepackage{dcolumn}% Align table columns on decimal point
\usepackage{bm}% bold math
\def\Cr23C6{{Cr$_{23}$C$_6$}}
\def\C6Cr23{{C$_6$Cr$_{23}$}}

\begin{document}

\title{Stability of Fe-based alloys with structure type \C6Cr23}

\author{M. Widom}
\author{M. Mihalkovi\v{c}}
 \altaffiliation[Also at ]{Institute of Physics, Slovak Academy of Sciences,
84228 Bratislava, Slovakia}
\affiliation{
Department of Physics\\
Carnegie Mellon University\\
Pittsburgh, PA  15213
}

\date{\today}

\begin{abstract}
Bulk metallic glass forms when liquid metal alloys solidify without
crystalization.  In the search for Iron-based bulk glass-forming
alloys of the metal-metalloid type (Fe-B- and Fe-C-based), crystals
based on the structural prototype \C6Cr23 often preempt the amorphous
phase.  Destabilizing this competing crystal structure could enhance
glass-formability.  We carry out first-principles total energy
calculations of enthalpy of formation to identify third elements that
can effectively destabilize \C6Cr23.  Yttrium appears optimal among
transition metals, and rare earths also are suitable.  Atomic size is
the dominant factor.
\end{abstract}

\maketitle

\section{\label{sec:Intro}Introduction}

Iron-based amorphous alloys are used in transformer cores, where their
low magnetic coercivity reduces energy loss.  Popular glass-forming
alloys are based on Fe together with metalloid elements such as B or
C.  Bulk Iron-based amorphous alloys could become important structural
materials, but optimal glass-forming compositions are not yet known.
Multicomponent alloys containing fourth row transition metals and rare
earths show promise~\cite{Poon03,Poon04,Lu04}.

We previously explored the quaternary B-Fe-Y-Zr phase
diagram~\cite{fezryb}, identifying stable and metastable crystal
phases and computing their enthalpies of formation.  This study
identified crystalline structures based on the \C6Cr23 prototype as
important competitors to glass formation.  It appeared that the
competition is more problematic in the case of B-Fe-Zr than in the
case of B-Fe-Y.  To ensure the optimal selection of alloy system, we
now carry out a systematic study of many candidate ``third elements''
and compare them with regard to stability of the \C6Cr23 structure.
We do this both for the case of B-Fe- and C-Fe-based alloys.  We show
that atomic size mismatch destabilizes the \C6Cr23 structure for
sufficiently large atoms such as Yttrium and rare
earths~\cite{Egami84,Miracle03,Poon04}.

\section{\label{sec:Methods}Methods}

Our {\em ab-initio} calculations use the program
VASP~\cite{VASP,VASP2} together with the projector-augmented wave
method, an all-electron generalization of the pseudopotential
approach~\cite{PAW,KJ_PAW}.  We employ the Perdew-Wang generalized
gradient approximation~\cite{PW91} (GGA) exchange-correlation
functional with the Vosko-Wilkes Nussair~\cite{VWN} spin
interpolation.  These choices give excellent results for bulk
elemental Fe~\cite{KJ_PAW}. GGA is needed instead of LDA is necessary
to properly reproduce magnetization and lattice
constants~\cite{Moroni}.  Our magnetic calculations are spin-polarized
(i.e. collinear magnetization) and are employed for any structure
containing 50\% Fe or higher.  All calculations for Carbon-based
binaries and ternaries are performed at a constant cutoff energy of
400 eV, the default for our Carbon potential.  All calculations for
Boron-based binaries and ternaries are performed at 320 eV, the
default for our Boron potential.  More details and discussion of
convergence, etc. are given in Ref.~\cite{fezryb}.  All the ab-initio
data on which this paper is based can be obtained on the WWW at
Ref.~\cite{alloy_home}.

\begin{figure*}[tbh]
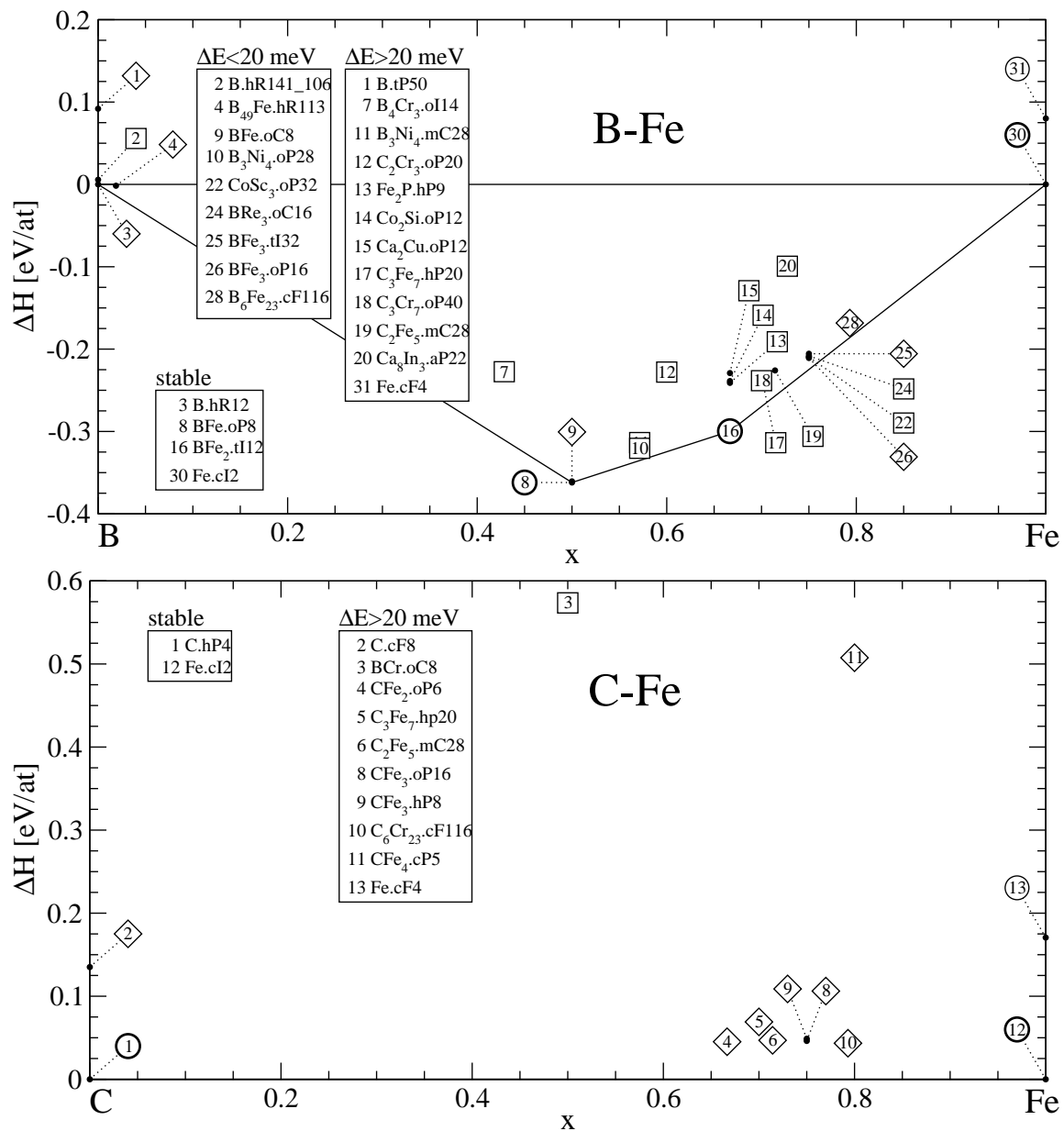

\includegraphics*[width=6in]{b-fe}
\includegraphics*[width=6in]{c-fe}
\caption{\label{fig:binaries} Enthalpies of formation and their convex
hulls for the B-Fe (top) and C-Fe (bottom) binary alloys. Notation: heavy
circles denote known low temperature phases, light circles denote
known high temperature phases, diamonds denote known metastable
phases, squares denote unreported structures.}
\end{figure*}

The composition space of an $N$-component alloy is a set of $N$
composition variables forming an $N-1$ dimensional simplex
(respectively, a point, line segment, triangle and tetrahedron for
$N=1, 2, 3, 4$).  Structural energies form a scatter-plot over this
simplex.  Stable low temperature phases lie on vertices of this
scatter-plot. Edges and facets of the convex hull represent
coexistence regions of the phases at adjoining vertices.

The tie-lines and tie-planes connecting all pure elements in their
lowest energy structures forms a useful reference for alloy energies.
The distance $\Delta H_{for}$ of an alloy energy from the tie-surface
joining pure elements is known as its enthalpy of formation.  It is an
enthalpy because volume relaxation means we work at fixed pressure,
$P=0$.  Strong compound formation is reflected in large negative
enthalpy of formation.  The value of $\Delta H$ is determined solely
by the cohesive energy of a given structure relative to the cohesive
energies of its constituent pure elements.

High temperature phases should lie above the convex hull, but be
sufficiently close that entropic effects (e.g. atomic vibrations,
vacancies or chemical substitution) can stabilize them.  Metastable
phases also should lie close to the convex hull, so that their free
energy is less than the liquid free energy at temperatures below
freezing.  Although $\Delta H_{for}$ is usually negative for high
temperature and metastable phases, their energy difference $\Delta E$
from the convex hull is small and positive.  $\Delta E$ measures the
thermodynamic driving force for decomposition into the appropriate
combination of stable phases.  In contrast to $\Delta H$, the value of
$\Delta E$ depends on the cohesive energies of other competing
structures.  Discovery of a new stable structure will increase the
assessed $\Delta E$ values of previously known structures.

\begin{figure*}[tbh]
\includegraphics*[width=6in]{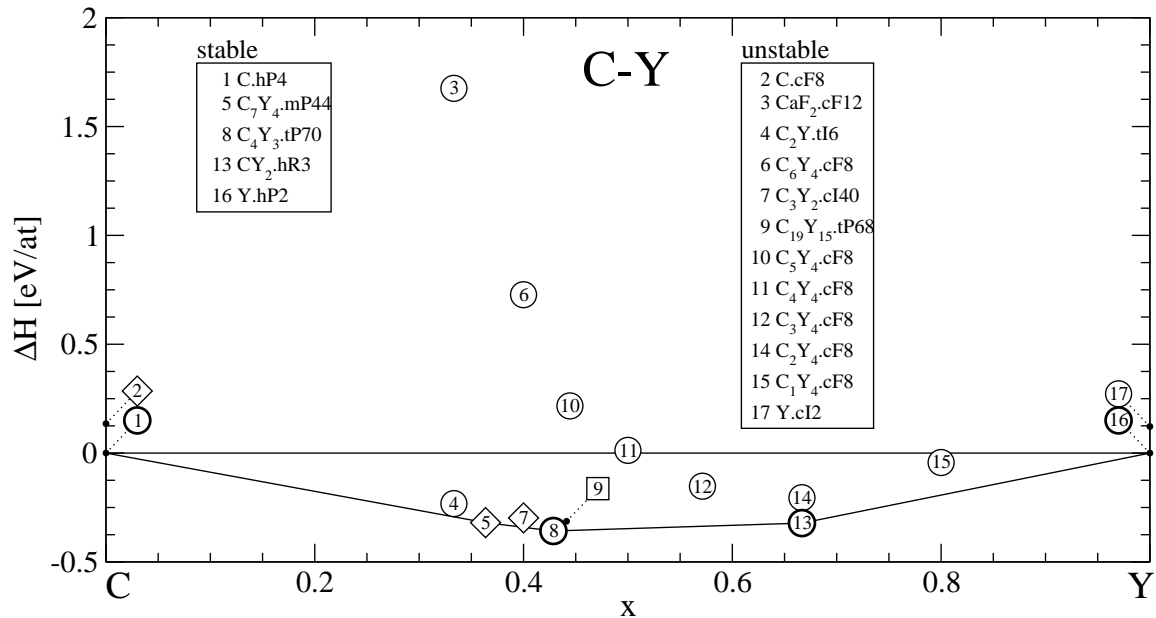}
\caption{\label{fig:CY} Enthalpies of formation and their convex
hull for the C-Y binary alloys. Notation as in Fig.~\ref{fig:binaries}}
\end{figure*}

In principle all possible combinations and arrangements of atoms
should be considered to ensure the optimal possibilities are found.
This is clearly impossible.  Rather, we choose plausible structures
for consideration by chemical substitution into known structures of
similar compounds~\cite{Pearson}.  We especially consider alloy
systems with similar atomic size ratios or other chemical properties.

Structures are denoted using their prototype names and Pearson
symbols.  For example, we will be interested in the \C6Cr23 prototype.
The element Cr will be replaced by Fe, and in some cases the element C
will be replaced by B.  The Pearson symbol for the \C6Cr23 prototype
is cF116, indicating cubic symmetry, face-centering, with 116 atoms
per unit cell.  The {\em primitive} cell of \C6Cr23.cF116 contains
116/4=29=23+6 atoms.

Using these methods, we built a database of structural energies.  For
a given $N$-component alloy system of interest we extract from our
database energies of structures containing all, of just some, of the
chosen elements.  We use a standard convex hull program
(qhull~\cite{qhull}) to identify stable structures and the coexistence
regions that connect them.  Based on the output of this program, we
calculate values of $\Delta H_{for}$ and $\Delta E$ for every
structure.  Numerical data for the compounds considered here, and more
than 1500 others, can be found at Ref.~\cite{alloy_home}.

\section{\label{sec:Results}Results}

\subsection{\label{sec:binary}Binaries}

Cohesive energies of binary B-Fe and C-Fe alloys are shown in
Fig.~\ref{fig:binaries}.  Our results for both alloy systems are in
perfect qualitative agreement with experiment~\cite{BinaryPD,PD_new}, because
all the known low temperature stable phases occur on the convex hulls
(notice C-Fe has no stable compounds) and all the know high
temperature, metastable and unknown hypothetical structures lie above
the hull.  It is impressive how sensitive the density functional
theory is to differences in chemical identity, with the many
distinctions between B-Fe and C-Fe all being faithfully reproduced.

Many metastable phases are known in the C-Fe binary system.  Two of
these, CFe$_4$.cP5 and CFe$_3$.hP8 are based on FCC Iron with Carbon
interstitials in, respectively, tetrahedral and octahedral sites.  The
CFe$_3$.oP16 structure is also an important metstable phase in B-Fe.
It can be considered as a strong distortion of the Fe$_3$Si.cF16
structure, caused by the small atomic sizes of B and C relative to Si.
Similarly, the energies of the interstitial cP5 and hP8 structures are
lower for C than for B because the Carbon atoms are smaller than
Boron.  The chief B-Fe structures are discussed further in
Ref.~\cite{fezryb}.

The lowest-lying metastable C-Fe binary compound has the structure of
\C6Cr23.  This crystal structure appears to be the most important
competitor to metallic glass formation, and further destabilizing it
is the goal of this work.

To evaluate stabilities of a ternary alloy system such as C-Fe-Y we
need to examine all three of its binary subsystems.  The Fe-Y diagram
was previously discussed in~\cite{fezryb} and need not be repeated
here.  The C-Y binary phase diagram is poorly known~\cite{C-Y,C-Y-upd},
with three phases ($\beta$-C$_{19}$Y$_{15}$ and
$\alpha,\beta$-C$_3$Y$_2$) of unknown structure.  The reported stable
low temperature tP68 structure of $\alpha$-C$_{19}$Y$_{15}$
(isostructural with C$_{19}$Sc$_{15}$.tP68) suffers from large
displacements during relaxation, and a final relaxed energy above the
convex hull.  We find instead that C$_4$Y$_3$.tP70~\cite{Pearson} is
stable at low temperatures.

The high temperature $\gamma$ phase probably takes the reported
Fe$_4$N.cF8 structure in the Y-rich limit.  This structure is based
upon an FCC lattice of Y atoms with C occupying octahedral
interstitial sites.  When these sites are fully occupied the unit cell
composition is C$_4$Y$_4$ and the structure becomes NaCl.  On the
C-rich side, we begin occupying tetrahedral interstitials and the
energy rapidly grows beyond values that are plausible for a high
temperature phase.  We also tested Y vacancies and found them even
higher in energy.  We believe the $\gamma$ phase actually terminates
at 50\% Carbon, rather than the reported 67\%.  The postulated
$\beta$-C$_2$Y.cF12 structure~\cite{C-Y} at 67\% Carbon (based on an
FCC lattice of Y with C fully occupying all tetrahedral interstitials)
is highly unstable and presumably incorrect.

\begin{figure*}
\includegraphics*[width=6in]{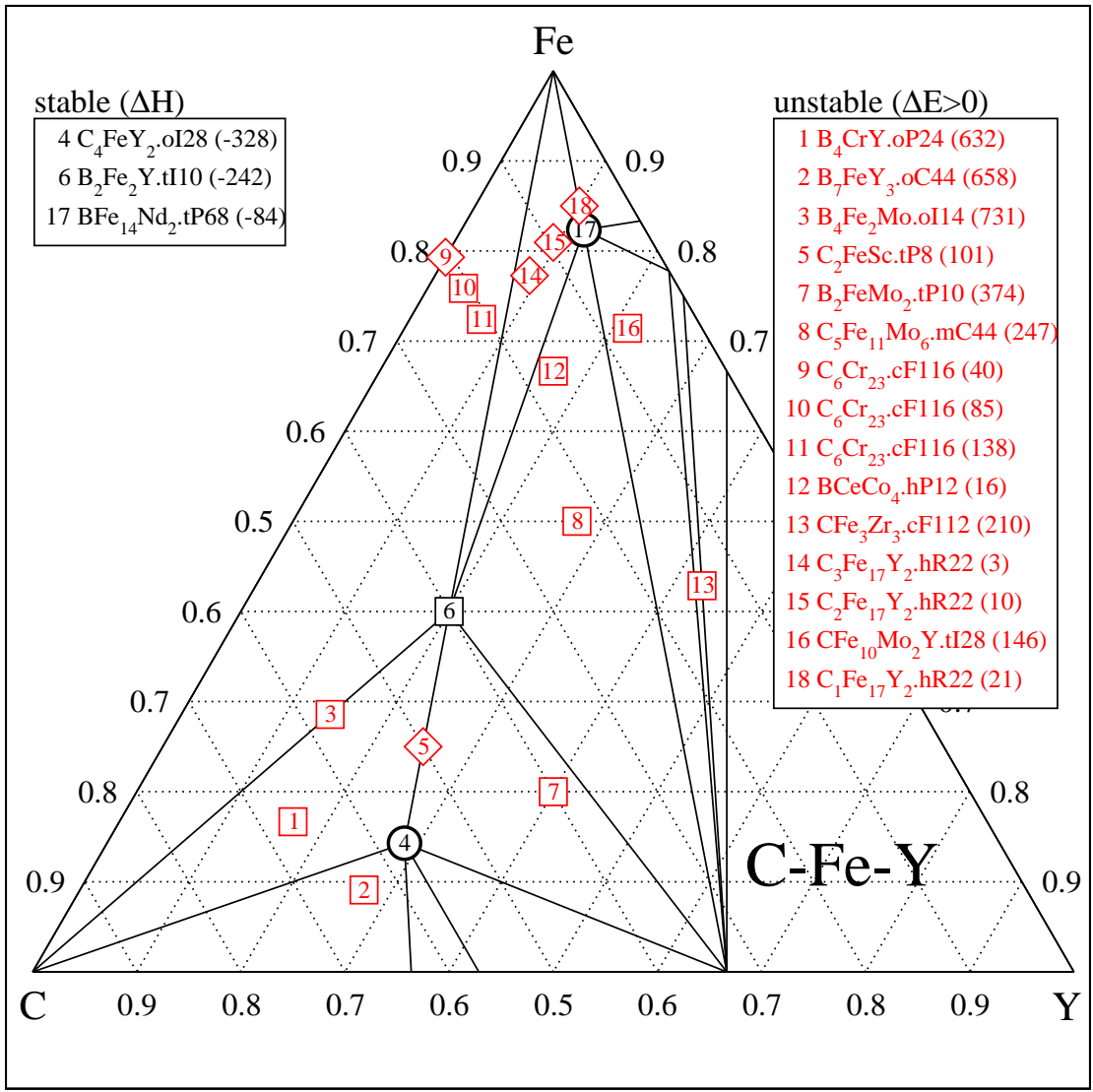}
\caption{\label{fig:tern} Energy diagram for ternary C-Fe-Y. Plotting
symbols as in Fig.~\ref{fig:binaries}.  Values in parenthesis indicate
enthalpy of formation $\Delta H_{for}$ for stable compounds, and
energies above the convex hull $\Delta E$ for metastable structures.}
\end{figure*}

\subsection{\label{sec:ternary}Ternaries}

Having verified the ability to reproduce these binary phase diagrams,
we now investigate ternary systems.  Fig~\ref{fig:tern} illustrates a
typical ternary energy diagram, in this case for the alloy C-Fe-Y.
This diagram confirms stability of previously known ternary phases and
also proposes B$_2$Fe$_2$Y.tI10 as stable, though this structure has
not previously been reported~\cite{TernaryPD}.  We also propose
C$_2$FeSc.tP8 as the probable structure of a previously reported
metastable phase of unknown structure.  Similar energy diagrams for
all ternaries discussed below may be found at Ref.~\cite{alloy_home}.

We now focus our attention on one specific structure, \C6Cr23, and
examine its stability for a variety of B-Fe- and C-Fe-based ternaries.
Although Chromium-based in the prototype structure, it is a well known
metastable phase in Iron-based alloys such as C-Fe-Mo and we predict
it is actually stable in B-Fe-Sc, B-Fe-Nb and B-Fe-Mo.

Our plan is to find elements that can mix with the binaries in the
liquid state but that will destabilize \C6Cr23 due to their large
atomic size.  The requirement of miscibility in the liquid state rules
out alkali metals and alkali earths.  The requirement of large atomic
size rules out middle and late transition metals.  We thus identify the
primary candidates as the early transition metals and the rare earths.

\begin{table}
\caption{\label{tab:sites}Site energies ($\Delta E$ (eV/atom)),
near-neighbor distances and Voronoi volumes for Y substitution on
different Wyckoff classes.}
\begin{tabular}{|r|r|r|r|r|}
\hline
site   & 4a & 8c & 32f & 48h \\
\hline
\hline
energy &  0.135  &  0.085  &  0.145   &  0.146   \\
$r_Y$  &  2.62   &  2.71   &  2.42    &  2.40  \\
$V$    & 20.1    & 24.5    & 23.7     &  22.4 \\
\hline
\end{tabular}
\end{table}

Our discussion begins with Yttrium, the first transition metal of row
4 in the periodic table.  The first question to address is the optimal
site for Y atoms.  As a large atom, it cannot enter as an
interstitial, nor can it substitute for Carbon.  There are thus 4
plausible sites, the Iron sites of Wyckoff classes 4a, 8c, 32f or 48h
(see Table~\ref{tab:sites}).

Table~\ref{tab:sites} evaluates the relaxed energies of a single 29
atom primitive cell of cF116 with one Y atom replacing one Fe atom,
respectively, on each distinct Fe site.  Evidently Y atoms favor
Wyckoff class 8c, most likely as a result of their large size.
Indeed, placing Y on site 8c results in both the largest Voronoi
volume $V$ for Y atoms and the largest near-neighbor distance $r_Y$.

We also checked that the 8c site is prefered for a {\em second} Y
substitution, after the first Y atom is already on 8c.  Since there
are only two 8c sites in a primitive cell (8/4=2) this corresponds to
complete filling of the 8c sites with Y atoms.  Double occupancy of
the 8c site by the large atom is consistent with the solved
crystallographic structure of B$_6$Co$_{21}$Zr$_2$, which is
isostructural with \C6Cr23.

The following energy study focuses on the case of double substitution
to fully occupy the 8c sites.  The energy cost for occupying both 8c
sites by Y is more than twice the cost for occupying just one, as can
be seen by comparing tables~\ref{tab:sites} and~\ref{tab:nrg}.  The
energies for 0, 1 and 2 Y-atom substitutions on site 8c in C6Fe23 are,
respectively, $\Delta E$=0.043, 0.085 and 0.141 eV/atom in a 29 atom
primitive cell.  So the first substitution costs $29\times
(0.085-0.043)=1.22$ eV, while the second costs $29\times
(0.141-0.085)=1.62$ eV.  Presumably this is because lattice strains
created by insertion of the first Y atom are accomodated partly by
shrinkage of the volume around the second 8c position.  When the
second Y atom is introduced this accomodation is no longer possible.

\begin{figure*}
\includegraphics[width=3in]{struct}
\includegraphics[width=3in]{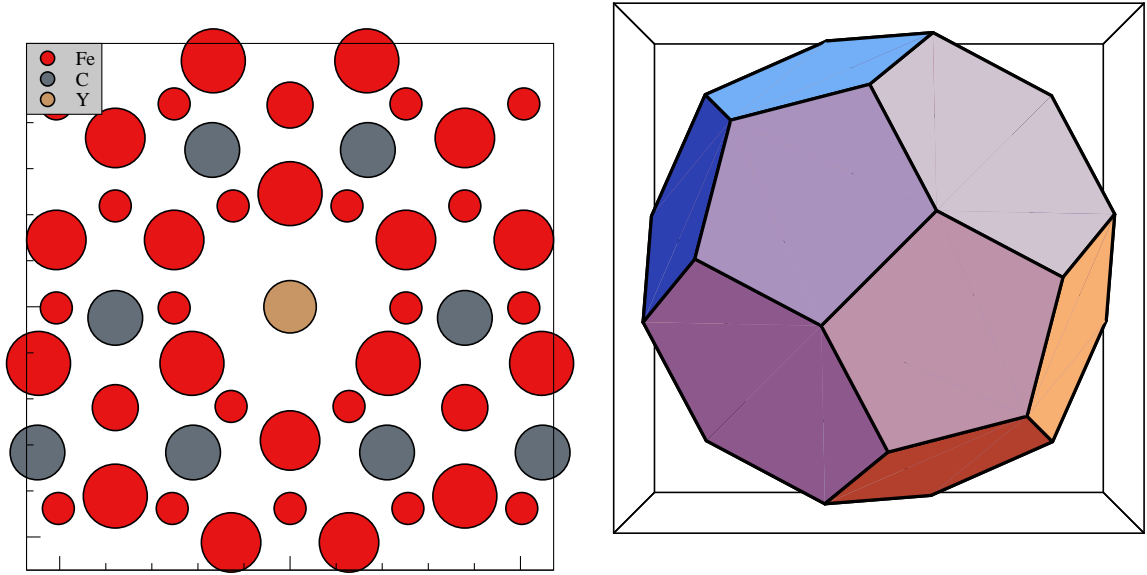}
\caption{\label{fig:struct} (left) Structure of C6Fe21Y2.cF116 viewed
along (1,1,1). Atom size indicates height. Hash marks are 1~\AA. (right)
Voronoi polyhedron of Y site, Wyckoff class 8c, viewed along (1,0,0).}
\end{figure*}

A picture of the cF116 structure at composition C$_6$Fe$_{21}$Y$_2$ is
shown in Fig.~\ref{fig:struct}.  The 8c sites have coordination number
$Z=16$, with only Fe atom neighbors.  The Voronoi polyhedron (Watson
and Bennett~\cite{WB} class (0,0,12,4)) has 12 pentagonal faces and 4
hexagonal faces, characteristic of local tetrahedral close packing.
The Voronoi polyhedron is illustrated in Fig.~\ref{fig:struct}.
However, the remainder of the structure does not exhibit tetrahedral
packing.

The Carbon sites, Wyckoff class 24e, have a Voronoi polyhedron (Watson
and Bennett class (0,8,0)) with 8 square faces and a highly nongeneric
four-fold vertex.  The eight neighbors of each Carbon atom are Iron
atoms at distances of 2.05 and 2.15~\AA, substantially greater than
the 1.79-1.99~\AA~ found in other metastable C-Fe structures.  This
loose binding of the Carbon atom probably explains why the \C6Cr23
structure is actually more favorable for B-Fe than for C-Fe, because
Boron is slightly larger.

Now we compare the stabilities with different choices of large atom.
Our data is presented in Table~\ref{tab:nrg} for several choices of
large atom.  In each case we verified that the 8c site is prefered.
We fully occupy the 8c site, with two large atoms per face-centered
primitive cell.  Values of $r_{T}$ are distances from large atom to
nearest neighbor transition metal in the fully relaxed structure.

Evidently, of the 4d transition metals, the largest atom, Yttrium, is
the most effective at destabilizing \C6Cr23.  Indeed, Zr, Nb and Mo
tend to {\em stabilize} it relative to the B-Fe and C-Fe binaries.
However, too large a Y concentration is dangerous for glass formation,
because of the low-energy metastable C$_3$Fe$_{17}$Y$_2$.hR22
structure (Y content 9.1\%) and the stable structure
CFe$_{14}$Y$_2$.tP68 (Y content 11.8\%) visible in
Fig.~\ref{fig:tern}.

We can also test whether the 4d row is an optimal choice by examining
other members of group IIIA (the first column of transition metals,
consisting of Sc, Y, La and Ac).  Clearly the 3d element Sc is too
small to destabilize \C6Cr23.  Lanthanum (5d) and Actinium (6d) are
both highly effective.  The lanthanide and actinide rare earth
elements, which behave chemically like Lanthanum and Actinium, should
likewise do well.  Of course, Actinides may present other difficulties
not considered here!

\begin{table}
\caption{\label{tab:nrg}Energies for complete TM substitution on site 8c.
the top row are B-Fe and C-Fe binaries for comparison.}
\begin{tabular}{|r||r|r|r||r|r|r|}
\hline
$^Z$Ch  &  \multicolumn{3}{c||}{B-Fe}  & \multicolumn{3}{c|}{C-Fe} \\
\hline
    & $\Delta E$ & $\Delta H$ & r$_T$ & $\Delta E$ & $\Delta H$ & r$_T$ \\
\hline
$^{26}$Fe &  0.018 & -0.168 & 2.40 &  0.043 &  0.043 & 2.42 \\
\hline
$^{39}$Y  &  0.074 & -0.163 & 2.61 &  0.141 &  0.059 & 2.61 \\
$^{40}$Zr &  0.016 & -0.253 & 2.54 &  0.025 & -0.037 & 2.54 \\
$^{41}$Nb & stable & -0.246 & 2.48 &  0.002 & -0.028 & 2.48 \\
$^{42}$Mo & stable & -0.212 & 2.42 &  0.030 &  0.022 & 2.41 \\
\hline
$^{21}$Sc & stable & -0.244 & 2.53 &  0.054 & -0.035 & 2.53 \\
$^{57}$La &  0.094 & -0.092 & 2.60 &  0.195 &  0.189 & 2.61 \\
$^{89}$Ac &  0.232 &  0.046 & 2.66 &  0.228 &  0.228 & 2.65 \\
\hline
\end{tabular}
\end{table}

\section{Conclusions}

We compared the effectiveness of different large-atom substitution as
a way to destabilize the \C6Cr23 structure of B-Fe and C-Fe.  The
first task was to identify the most energetically favorable
substitution site for the large atom.  We found, in every case
considered that Wyckoff site 8c (a locally tetrahedrally close-packed
site) was most favorable.  We showed that among the 4d transition
metals Yttrium was the only element capable of destabilizing the
\C6Cr23 structure relative to the binary alloy.  Comparing group IIIA
elements, we found that Sc was inadequate but Y, La and Ac are fine.

Clearly atomic size is a crucial consideration for destabilizing this
structure.  Indeed, the 8c site prefers atoms slightly larger than Fe,
so replacement with Mo, Nb or Sc can actually {\em stabilize} cF116.
We conclude that Yttrium and rare earth elements can enhance glass
formation.  However, too large an Yttrium concentration (9\% or more)
can lead to formation of other competing crystal phases such as
C$_3$Fe$_{17}$Y$_2$.hR22 or BFe$_{14}$Nd$_2$.tP68.  Alkali earth
elements such as Ca are also highly effective at destabilizing
\C6Cr23, but they do not mix well with Fe-based alloys, at least at
atmospheric pressure.

\begin{acknowledgments}
We wish to acknowledge useful discussions with Yang Wang, Don
Nicholson, Joe Poon and Gary Shiflet.  This work was supported by
DARPA/ONR Grant N00014-01-1-0961.  Portions of the calculations were
performed at the Pittsburgh Supercomputer Center.
\end{acknowledgments}

\newpage

\bibliography{stability}

\end{document}